\documentclass[aps,prl,preprint,tightenlines,superscriptaddress,showpacs,byrevtex]{revtex4}
\usepackage{graphicx} 
\usepackage{dcolumn}  

\graphicspath{{ps}}

\newcommand{\BB}{B\bar{B}}
\newcommand{\BF}{\mathcal{B}}

\newcommand{\bsgamma}{b \to s\gamma}
\newcommand{\CP}{\mathit{CP}}

\newcommand{\coshel}{\cos\theta_\mathrm{hel}}
\newcommand{\DE}{\Delta E}
\newcommand{\Ebeam}{E_{\mathrm{beam}}^{*}}
\newcommand{\Egamma}{E_\gamma^*}

\newcommand{\fbi}{\mathrm{fb}^{-1}}
\newcommand{\GeV}{\mathrm{GeV}}
\newcommand{\Ktwost}{K_2^*(1430)}
\newcommand{\Kthreest}{K_3^*(1780)}

\newcommand{\Mbc}{M_{\mathrm{bc}}}
\newcommand{\MeV}{\mathrm{MeV}}
\newcommand{\MKeta}{M_{K\eta}}
\newcommand{\MXs}{M_{X_s}}
\newcommand{\KS}{K^0_S}
\newcommand{\qq}{q\bar{q}}

\newcommand{\PM}[2]{{\,}^{+#1}_{-#2}}

\newcommand{\LumONRES}{140}
\newcommand{\LumOFFRES}{15}
\newcommand{\NBBmillunit}{152}

\newcommand{\BFbsgammaPDG}{(3.3 \pm 0.4) \times 10^{-4}}

\newcommand{\effKID}{90\%}
\newcommand{\fakeKID}{9\%}
\newcommand{\effPIID}{98\%}

\newcommand{\MCeffLRsig}{44\%}
\newcommand{\MCrejLRqq}{98\%}

\newcommand{\YieldCharged}{35.5 \PM{10.1}{9.4} \PM{5.6}{3.9}}
\newcommand{\YieldNeutral}{9.7 \PM{5.1}{4.3} \PM{1.6}{1.5}}
\newcommand{\YieldTotal}{45.0 \PM{11.2}{10.4} \PM{7.4}{5.1}}
\newcommand{\YieldTc}{10.5 \PM{5.6}{4.8} \PM{2.8}{2.4}}
\newcommand{\YieldTn}{4.2 \PM{3.2}{2.4} \PM{0.7}{0.9}}
\newcommand{\YieldTt}{15.0 \PM{6.3}{5.5} \PM{3.3}{3.1}}
\newcommand{\YieldULTc}{21.5}
\newcommand{\YieldULTn}{10.0}
\newcommand{\YieldULTt}{27.3}

\newcommand{\YieldChargedNosyst}{35.5 \PM{10.1}{9.4}}
\newcommand{\YieldNeutralNosyst}{9.7 \PM{5.1}{4.3}}
\newcommand{\YieldTotalNosyst}{45.0 \PM{11.2}{10.4}}

\newcommand{\SNFstatCharged}{4.3}
\newcommand{\SNFstatNeutral}{2.5}
\newcommand{\SNFstatTotal}{5.0}

\newcommand{\EFFc}{3.40 \pm 0.20}
\newcommand{\EFFn}{0.88 \pm 0.12}
\newcommand{\EFFt}{4.28 \pm 0.25}
\newcommand{\EFFTc}{0.77 \pm 0.34}
\newcommand{\EFFTn}{0.17 \pm 0.08}
\newcommand{\EFFTt}{0.94 \pm 0.41}

\newcommand{\SYSTphotonSTR}{2.8\%}
\newcommand{\SYSTkaonidSTR}{0.8\%}
\newcommand{\SYSTpionidSTR}{0.5\%}
\newcommand{\SYSTksSTR}{4.5\%}
\newcommand{\SYSTpizeroSTR}{1.5\%}
\newcommand{\SYSTetaSTR}{2.0\%}
\newcommand{\SYSTlrsystCstr}{3.2\%}
\newcommand{\SYSTlrsystNstr}{9.0\%}
\newcommand{\SYSTketamassCstr}{2.1\%}
\newcommand{\SYSTketamassNstr}{8.4\%}
\newcommand{\SYSTsubbfgSTR}{0.7\%}
\newcommand{\SYSTsubbfpSTR}{1.8\%}

\newcommand{\BFketagammaZc}{6.9 \PM{2.0}{1.8} \PM{1.2}{0.9}}
\newcommand{\BFketagammaZn}{7.3 \PM{3.8}{3.2} \PM{1.6}{1.5}}
\newcommand{\BFketagammaZt}{6.9 \PM{1.7}{1.6} \PM{1.3}{1.0}}
\newcommand{\BFkthreestgammaULc}{33}
\newcommand{\BFkthreestgammaULn}{72}
\newcommand{\BFkthreestgammaULt}{34}

\newcommand{\BFketagammaZtSTR}{( 6.9 \PM{1.7}{1.6} \mbox{(stat)}%
 \PM{1.3}{1.0} \mbox{(syst)} ) \times 10^{-6}}

\begin{document}


\preprint{\vbox{ \hbox{   }
                 \hbox{BELLE-CONF-0412}
                 \hbox{ICHEP04 11-0657}
}}

\title{ \quad\\[0.5cm]%
Observation of {\protect\boldmath$B \to K\eta\gamma$}}

\affiliation{Aomori University, Aomori}
\affiliation{Budker Institute of Nuclear Physics, Novosibirsk}
\affiliation{Chiba University, Chiba}
\affiliation{Chonnam National University, Kwangju}
\affiliation{Chuo University, Tokyo}
\affiliation{University of Cincinnati, Cincinnati, Ohio 45221}
\affiliation{University of Frankfurt, Frankfurt}
\affiliation{Gyeongsang National University, Chinju}
\affiliation{University of Hawaii, Honolulu, Hawaii 96822}
\affiliation{High Energy Accelerator Research Organization (KEK), Tsukuba}
\affiliation{Hiroshima Institute of Technology, Hiroshima}
\affiliation{Institute of High Energy Physics, Chinese Academy of Sciences, Beijing}
\affiliation{Institute of High Energy Physics, Vienna}
\affiliation{Institute for Theoretical and Experimental Physics, Moscow}
\affiliation{J. Stefan Institute, Ljubljana}
\affiliation{Kanagawa University, Yokohama}
\affiliation{Korea University, Seoul}
\affiliation{Kyoto University, Kyoto}
\affiliation{Kyungpook National University, Taegu}
\affiliation{Swiss Federal Institute of Technology of Lausanne, EPFL, Lausanne}
\affiliation{University of Ljubljana, Ljubljana}
\affiliation{University of Maribor, Maribor}
\affiliation{University of Melbourne, Victoria}
\affiliation{Nagoya University, Nagoya}
\affiliation{Nara Women's University, Nara}
\affiliation{National Central University, Chung-li}
\affiliation{National Kaohsiung Normal University, Kaohsiung}
\affiliation{National United University, Miao Li}
\affiliation{Department of Physics, National Taiwan University, Taipei}
\affiliation{H. Niewodniczanski Institute of Nuclear Physics, Krakow}
\affiliation{Nihon Dental College, Niigata}
\affiliation{Niigata University, Niigata}
\affiliation{Osaka City University, Osaka}
\affiliation{Osaka University, Osaka}
\affiliation{Panjab University, Chandigarh}
\affiliation{Peking University, Beijing}
\affiliation{Princeton University, Princeton, New Jersey 08545}
\affiliation{RIKEN BNL Research Center, Upton, New York 11973}
\affiliation{Saga University, Saga}
\affiliation{University of Science and Technology of China, Hefei}
\affiliation{Seoul National University, Seoul}
\affiliation{Sungkyunkwan University, Suwon}
\affiliation{University of Sydney, Sydney NSW}
\affiliation{Tata Institute of Fundamental Research, Bombay}
\affiliation{Toho University, Funabashi}
\affiliation{Tohoku Gakuin University, Tagajo}
\affiliation{Tohoku University, Sendai}
\affiliation{Department of Physics, University of Tokyo, Tokyo}
\affiliation{Tokyo Institute of Technology, Tokyo}
\affiliation{Tokyo Metropolitan University, Tokyo}
\affiliation{Tokyo University of Agriculture and Technology, Tokyo}
\affiliation{Toyama National College of Maritime Technology, Toyama}
\affiliation{University of Tsukuba, Tsukuba}
\affiliation{Utkal University, Bhubaneswer}
\affiliation{Virginia Polytechnic Institute and State University, Blacksburg, Virginia 24061}
\affiliation{Yonsei University, Seoul}
  \author{K.~Abe}\affiliation{High Energy Accelerator Research Organization (KEK), Tsukuba} 
  \author{K.~Abe}\affiliation{Tohoku Gakuin University, Tagajo} 
  \author{N.~Abe}\affiliation{Tokyo Institute of Technology, Tokyo} 
  \author{I.~Adachi}\affiliation{High Energy Accelerator Research Organization (KEK), Tsukuba} 
  \author{H.~Aihara}\affiliation{Department of Physics, University of Tokyo, Tokyo} 
  \author{M.~Akatsu}\affiliation{Nagoya University, Nagoya} 
  \author{Y.~Asano}\affiliation{University of Tsukuba, Tsukuba} 
  \author{T.~Aso}\affiliation{Toyama National College of Maritime Technology, Toyama} 
  \author{V.~Aulchenko}\affiliation{Budker Institute of Nuclear Physics, Novosibirsk} 
  \author{T.~Aushev}\affiliation{Institute for Theoretical and Experimental Physics, Moscow} 
% ...
\collaboration{The Belle Collaboration}

\noaffiliation

\begin{abstract}
 We report measurements of radiative $B$ decays with $K\eta\gamma$
 final states, using a data sample of $\LumONRES~\fbi$ recorded at the
 $\Upsilon(4S)$ resonance with the Belle detector at the KEKB $e^+e^-$
 asymmetric energy collider.
 We observe $B \to K\eta\gamma$ for the first time
 with a branching fraction of $\BFketagammaZtSTR$ for
 $\MKeta < 2.4~\GeV/c^2$.
 We also set an upper limit on $B \to \Kthreest\gamma$.
\end{abstract}

\pacs{13.20.He, 14.40.Nd}

\maketitle

\tighten

{\renewcommand{\thefootnote}{\fnsymbol{footnote}}}
\setcounter{footnote}{0}


Radiative $B$ decays,
which proceed mainly through the $\bsgamma$ process~\cite{footnote:charge},
have played an important role in the search for physics beyond
the Standard Model (SM).
Although the inclusive branching fraction has been measured
to be $\BFbsgammaPDG$~\cite{Eidelman:2004wy},
we know little about its constituents.
So far, measured exclusive final states such as
$K^*(892)\gamma$~\cite{Coan:1999kh,Aubert:2001me-Nakao:2004th},
$\Ktwost\gamma$~\cite{Coan:1999kh,Nishida:2002me},
$K\pi\pi\gamma$~\cite{Nishida:2002me} and
$B \to K\phi\gamma$~\cite{Drutskoy:2003xh}
only explain one third of the inclusive rate.
Detailed knowledge of exclusive final states
reduces the theoretical uncertainty in the measurement
of the inclusive branching fraction
using the pseudo-reconstruction technique,
as well as in the measurement of $B \to X_s\ell^+\ell^-$~\cite{Kaneko:2002mr}.
In this analysis, the decay mode $B \to K\eta\gamma$
is studied for the first time.
In addition to improving the understanding of $\bsgamma$ final states,
$B^0 \to \KS\eta\gamma$ can be used
to study
time-dependent $\CP$ asymmetry~\cite{Atwood:1997zr},
which is sensitive to physics beyond the SM.
The mode $B \to K\eta\gamma$ can also be used to search for
$B \to \Kthreest\gamma$ via $\Kthreest \to K\eta$ decay.

The analysis is based on
$\LumONRES~\fbi$ of data taken at the $\Upsilon(4S)$ resonance
(on-resonance) and
$\LumOFFRES~\fbi$ at an energy $60~\MeV$ below the resonance
(off-resonance),
which were recorded by the Belle detector~\cite{Mori:2000cg}
at the KEKB asymmetric $e^+e^-$ collider
($3.5~\GeV$ on $8~\GeV$)~\cite{KEKB:NIM}.
The on-resonance data corresponds to $\NBBmillunit$ million $\BB$ events.
The Belle detector
has a three-layer silicon vertex detector,
a 50-layer central drift chamber (CDC), an array of aerogel
Cherenkov counters (ACC), time-of-flight scintillation counters (TOF)
and an electromagnetic calorimeter of CsI(Tl) crystals (ECL) located inside
a superconducting solenoid coil that provides a 1.5 T magnetic field.
An instrumented iron flux-return for $K_L$/$\mu$ detection
is located outside the coil.

We reconstruct $B^+ \to K^+\eta\gamma$ and $B^0 \to \KS\eta\gamma$
via $\eta \to \gamma\gamma$ and $\eta \to \pi^+\pi^-\pi^0$.
All the charged tracks used in the reconstruction
(except charged pions from $\KS$)
are required to have an impact parameter
within $\pm 5 \mathrm{~cm}$
of the interaction point along the positron beam axis
and within $0.5 \mathrm{~cm}$ in the transverse plane.
Primary charged kaons are also required to have
a momentum in the $e^+e^-$ center-of-mass (CM) frame that is greater
than $100~\MeV/c$.
In order to identify kaon and pion candidates,
we use a likelihood ratio based on the light yield in the ACC,
the TOF information and the specific ionization measurements in the CDC.
For the selection applied on the likelihood ratio,
we obtain an efficiency
(pion misidentification probability) of $\effKID$ ($\fakeKID$)
for charged kaon candidates,
and an efficiency
(kaon misidentification probability) of $\effPIID$ ($\fakeKID$)
for charged pion candidates.

$\KS$ candidates are formed from $\pi^+\pi^-$ combinations
whose invariant mass is within $8~\MeV/c^2$
of the nominal $\KS$ mass.
The two pions are required to have a common vertex that
is displaced from the interaction point.
The $\KS$ momentum direction is also required to be
consistent with the $\KS$ flight direction.
Neutral pion candidates are formed from pairs of photons
that have an invariant mass within $16~\MeV/c^2$ of
the nominal $\pi^0$ mass and an energy greater than $100~\MeV$
in the CM frame.
Each photon is required to have an energy greater than $50~\MeV$.
A mass constrained fit is then performed to obtain the $\pi^0$ momentum.

$\eta$ candidates are reconstructed via $\eta \to \gamma\gamma$
or $\eta \to \pi^+\pi^-\pi^0$.
For $\eta \to \gamma\gamma$, we require that
the invariant mass of the two photons satisfy
$0.515~\GeV/c^2 < M_{\gamma\gamma} < 0.570~\GeV/c^2$
and that each photon have an energy greater than $50~\MeV$.
We also require that the $\eta$ decay helicity angle $\theta_\mathrm{hel}$
satisfies $|\coshel| < 0.9$.
A mass constrained fit is then performed to obtain the $\eta$ momentum.
For $\eta \to \pi^+\pi^-\pi^0$,
we apply a selection on the $M_{\pi^+\pi^-\pi^0}$ invariant mass,
$0.532~\GeV/c^2 < M_{\pi^+\pi^-\pi^0} < 0.562~\GeV/c^2$.

We combine a charged or neutral kaon with an $\eta$
to form a $K\eta$ system with invariant mass less than $2.4~\GeV/c^2$.
We then reconstruct $B$ meson candidates from the $K\eta$ system
and the highest energy photon
with a CM energy between $1.8~\GeV$ and $3.4~\GeV$
within the acceptance of the barrel ECL
($33^\circ<\theta_\gamma<128^\circ$, where $\theta_\gamma$
is the polar angle of the photon in the laboratory frame).
The photon candidate is required
to be consistent with an isolated electromagnetic shower,
i.e.\ 95\% of its energy should be concentrated in
an array of $3 \times 3$ crystals
and no charged tracks should be associated with it.
In order to reduce the background from
decays of $\pi^0$ and $\eta$ mesons,
we combine the photon candidate
with all other photons in the event
and reject the event if the invariant mass of any pair is
within $18~\MeV/c^2$ ($32~\MeV/c^2$)
of the nominal $\pi^0$ ($\eta$) mass
(this condition is referred to as the $\pi^0/\eta$ veto).

We use two independent kinematic variables for the $B$ reconstruction:
the beam-energy constrained mass
$\Mbc \equiv \sqrt{\left(\Ebeam/c^2\right)^2
  - (|\vec{p}_{K\eta}^{\,*}+\vec{p}_\gamma^{\,*}|/c)^2}$
and
$\DE \equiv E_{K\eta}^* + \Egamma - \Ebeam$,
where $\Ebeam$ is the beam energy,
and $\vec{p}_\gamma^{\,*}$, $\Egamma$,
$\vec{p}_{K\eta}^{\,*}$, $E_{K\eta}^*$ are
the momenta and energies of the photon
and the $K\eta$ system, respectively, calculated in the CM frame.
In the $\Mbc$ calculation, the photon momentum is rescaled
so that $|\vec{p}_\gamma^{\,*}|=(\Ebeam-E_{K\eta}^*)/c$
is satisfied.
We require $\Mbc > 5.2~\GeV/c^2$ and $-150~\MeV < \DE < 80~\MeV$.
We define the signal region to be $\Mbc > 5.27~\GeV/c^2$.
In the case that multiple candidates are found in the same event,
we take the candidate that has the $\eta$ mass closest to the nominal mass
(and smallest $|\DE|$)
after applying the background suppression described later.

The largest source of background originates from
continuum $e^+e^- \to \qq$ ($q = u,d,s,c$) production
including contributions from
initial state radiation ($e^+e^-\to q\bar{q}\gamma$).
In order to suppress this background,
we use the likelihood ratio (LR) described in Ref.~\cite{Nishida:2002me},
which utilizes the information from
a Fisher discriminant~\cite{Fisher:1936et}
formed from six modified Fox-Wolfram moments~\cite{Fox:1978vu}
and the cosine of the angle between
the $B$ meson flight direction and the beam axis.
The LR selection
retains $\MCeffLRsig$ of the signal,
rejecting $\MCrejLRqq$ of the continuum background.

In order to extract the signal yield, we perform
a binned likelihood fit to the $\Mbc$ distribution.
The $\Mbc$ distribution of the signal component is modeled
by a Crystal Ball line shape~\cite{Crystal-Ball},
where the parameters are determined from the signal MC
and calibrated by $B \to D\pi$ decays, as described below.
The $\Mbc$ distribution of the continuum background is
modeled by an ARGUS function~\cite{Albrecht:1990am}
whose shape is determined from the off-resonance data.
Here, the LR selection is not applied to the off-resonance data
in order to compensate for the limited amount of data in that sample.
The possible bias due to this is taken as systematic error to the fitted
yield.
Background from $B$ decays is divided into two components, which
we refer to as $\BB$ background and rare $B$ background in this paper.
The former comprises $B$ decays through $b \to c$ transitions including
color-suppressed $B$ decays such as $B^0 \to \bar{D}^0\pi^0$,
and the latter covers charmless $B$ decays through $b \to s$ and $b \to u$
transitions. Each of them is modeled by another ARGUS function.
The shape of the distributions is determined by
a corresponding Monte Carlo (MC) sample.
In order to study the contamination from other $\bsgamma$ decays,
we examine a $B \to K^*(892)\gamma$ MC sample
and an inclusive $\bsgamma$ MC sample that is
modeled as an equal mixture of $s\bar{d}$ and $s\bar{u}$
quark pairs and is hadronized using JETSET~\cite{Sjostrand:1994yb},
where the $\MXs$ spectrum is fitted to the model of Kagan and
Neubert~\cite{Kagan:1998ym}.
We find that the feed-down from other $\bsgamma$ decays is not negligible,
and its $\Mbc$ distribution is also modeled by an ARGUS function.

\begin{figure}
 \begin{center}
  \begin{tabular}{cc}
   (a) $B^+ \to K^+ \eta\gamma$ & (b) $B^0 \to \KS \eta\gamma$ \\
   \includegraphics[scale=0.68]{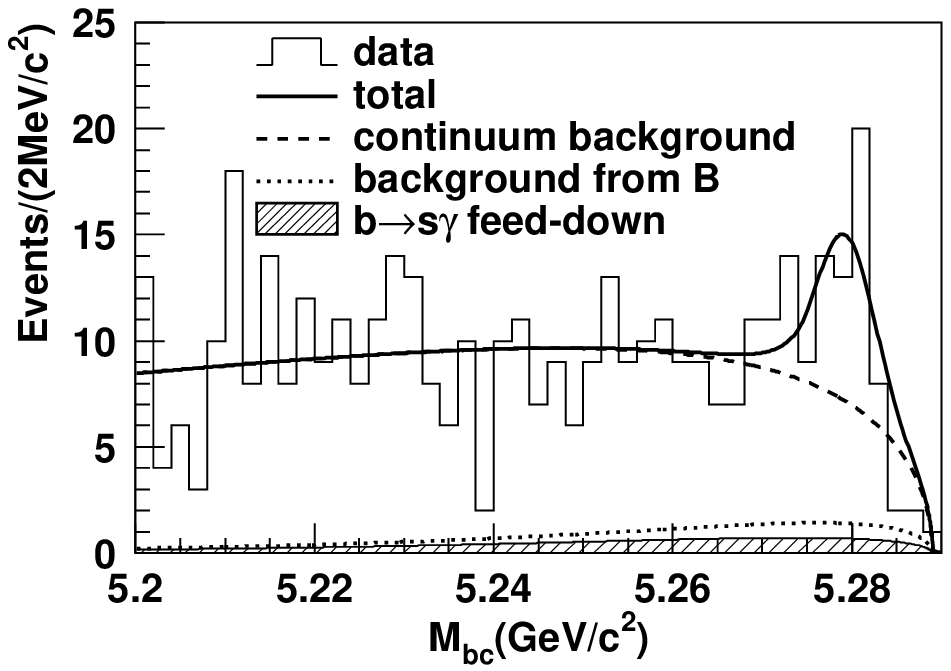} &
   \includegraphics[scale=0.68]{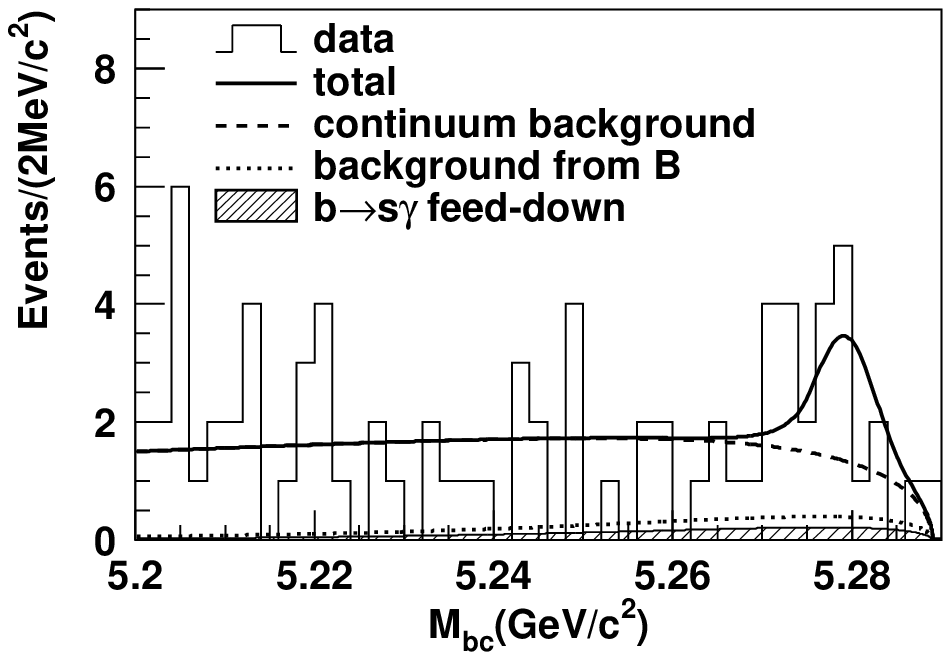} \\
   (c) $B \to K\eta\gamma$ & (d) $1.6~\GeV/c^2 < \MKeta < 1.95~\GeV/c^2$ \\
   \includegraphics[scale=0.68]{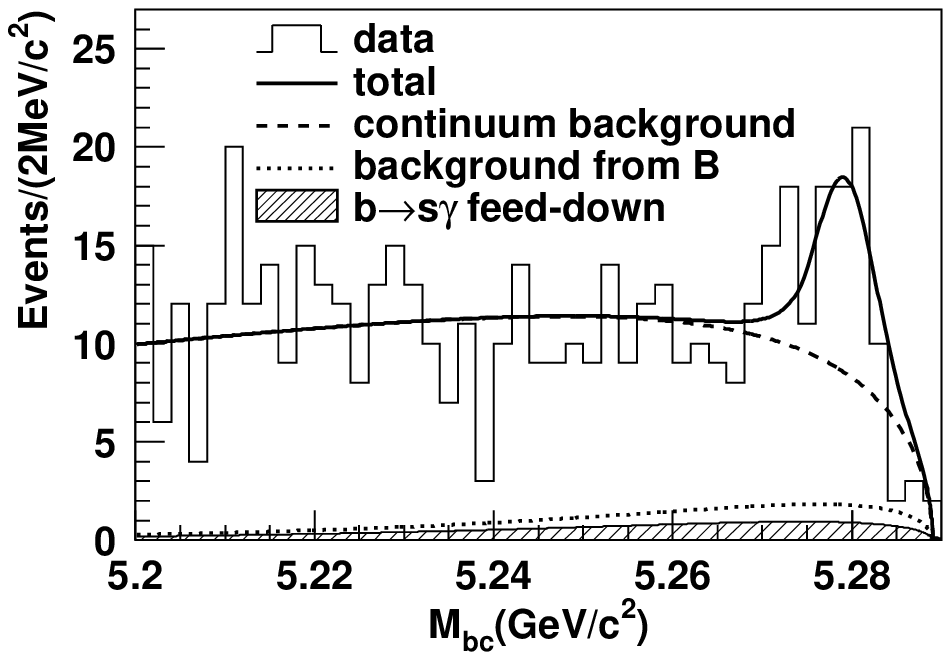} &
   \includegraphics[scale=0.68]{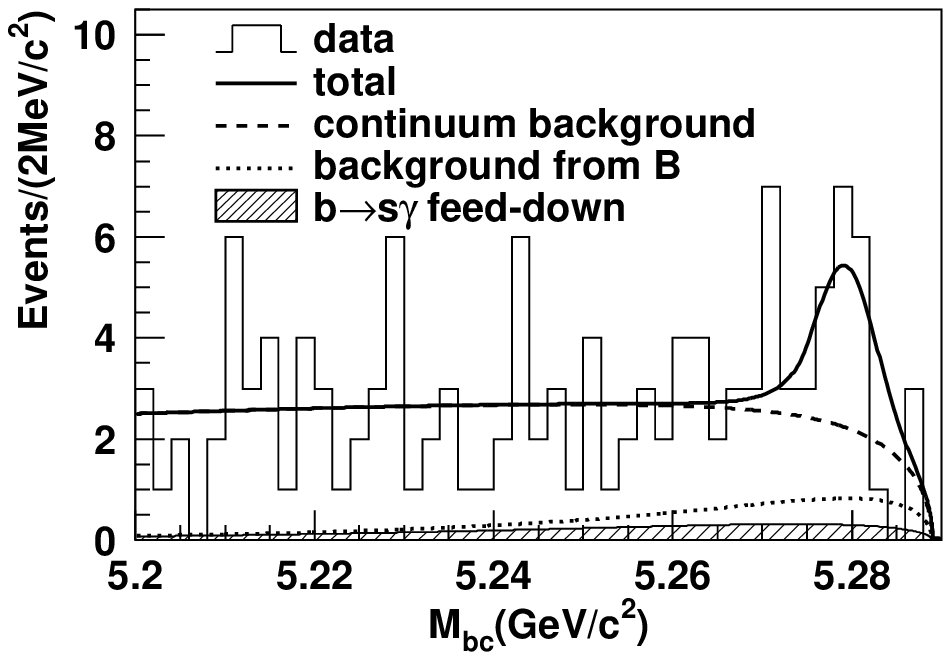} \\
  \end{tabular}
  \caption{\label{fig:mbcfit}%
  $\Mbc$ distributions for (a) $B^+ \to K^+ \eta\gamma$,
  (b) $B^0 \to \KS \eta\gamma$,
  (c) $B \to K\eta\gamma$
  (combined $B^+ \to K^+ \eta\gamma$ and $B^0 \to \KS \eta\gamma$),
  and (d) $B \to K\eta\gamma$ with the mass range for $B \to \Kthreest\gamma$.
  Fit results are overlaid.}
 \end{center}
\end{figure}

Figures~\ref{fig:mbcfit} (a)-(c) show the $\Mbc$ distributions
for $B^+ \to K^+ \eta\gamma$, $B^0 \to \KS \eta\gamma$
and combined $B \to K\eta\gamma$.
The distributions are fitted with signal, continuum,
$\BB$, rare $B$ background and the $\bsgamma$ feed-down components.
In the fit, the normalization of $\BB$, rare $B$ and $\bsgamma$
are fixed according to the luminosity and $\bsgamma$ branching fraction,
while the normalization of the continuum component is allowed to float.
We find the signal yields to be $\YieldChargedNosyst$,
$\YieldNeutralNosyst$ and $\YieldTotalNosyst$ with
statistical significances of
$\SNFstatCharged\sigma$, $\SNFstatNeutral\sigma$ and $\SNFstatTotal\sigma$,
for the charged, neutral and combined modes, respectively.
Here, the significance is defined as
$\sqrt{- 2 \ln ( {\cal L}(0) / {\cal L}_{\mathrm{max}} )}$, where
${\cal L}_{\mathrm{max}}$ is the maximum of the likelihood
and ${\cal L}(0)$ is the likelihood
for zero signal yield.

\begin{figure}
 \begin{center}
  \includegraphics[scale=0.7]{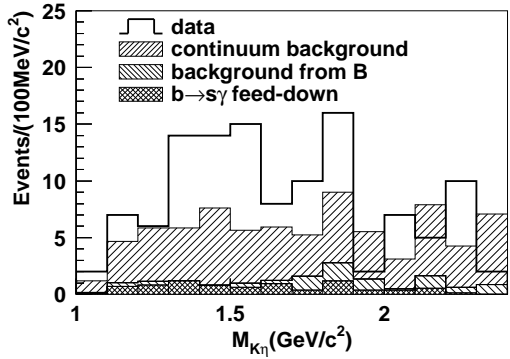}
  \caption{\label{fig:xsmass-keta}%
  $K\eta$ invariant mass distribution for events in the signal regions
  for $B \to K\eta\gamma$
  }
 \end{center}
\end{figure}

The $K\eta$ invariant mass distribution for events
inside the signal region is shown in Fig.~\ref{fig:xsmass-keta}.
Here, the background distributions are obtained from the
corresponding MC samples, and are normalized using the fit result.
We find that signal excess
is concentrated between $1.3~\GeV/c^2$ and $1.9~\GeV/c^2$.
Therefore, our selection $\MKeta < 2.4~\GeV/c^2$
is expected to include most of the $B \to K\eta\gamma$ signal.
We do not see any clear resonant structure in the $\MKeta$ distribution.

The systematic error on the signal yield due to the fitting procedure
is estimated by varying the value of each fixed parameter
by $\pm 1\sigma$ and extracting the new signal yield for each case.
The difference between the background shape for the continuum MC with
and without the LR selection is taken as an additional error to
the continuum background shape.
We set the normalization of either the $\BB$ or rare $B$
backgrounds to zero and to twice its nominal value
to account for its uncertainty.
The changes of the yields for each procedure
are added in quadrature, and are regarded as the systematic error
on the signal yield.

The signal reconstruction efficiency is estimated using
the MC simulation and is corrected for discrepancies
between data and MC using control samples.
We find that the efficiency is almost independent of the $K\eta$
invariant mass. Table~\ref{tab:summary} shows
the signal efficiencies and the branching fractions
for each $B \to K\eta\gamma$ mode.
Here, we assume an equal production rate for
$B^0\bar{B}^0$ and $B^+B^-$.
The error includes the following systematic uncertainties:
photon detection ($\SYSTphotonSTR$),
tracking ($1\%$ track),
kaon identification ($\SYSTkaonidSTR$),
pion identification ($\SYSTpionidSTR$ per pion),
$\KS$ detection ($\SYSTksSTR$),
$\pi^0$ detection ($\SYSTpizeroSTR$),
$\eta$ detection in $\eta \to \gamma\gamma$ mode ($\SYSTetaSTR$),
$\pi^0/\eta$ veto and LR ($\SYSTlrsystCstr$ and $\SYSTlrsystNstr$ for
charged and neutral modes, respectively),
possible $K\eta$ mass dependence of the efficiency
($\SYSTketamassCstr$ and $\SYSTketamassNstr$ for
charged and neutral modes, respectively), and
uncertainty in the $\eta$ branching fraction
($\SYSTsubbfgSTR$ for $\eta \to \gamma\gamma$
and $\SYSTsubbfpSTR$ for $\eta \to \pi^+\pi^-\pi^0$).
The systematic errors from the $\pi^0/\eta$ veto and LR requirement
are estimated
using the $B^- \to D^0(\to K^-\pi^+\pi^0)\pi^-$ and
$B^0 \to D^-(\to \KS\pi^-\pi^0)\pi^+$ as a control sample,
treating the primary pion
as a high energy photon. This sample is also used
to obtain the $\Mbc$ shape of the signal component.

We also search for the decay $B \to \Kthreest\gamma$ by applying
the additional requirement $1.60~\GeV/c^2 < \MKeta < 1.95~\GeV/c^2$.
The fits to the $\Mbc$ distributions yield
$\YieldTc$, $\YieldTn$ and $\YieldTt$ events
for charged, neutral and combined modes,
respectively~\cite{footnote:error}.
The $\Mbc$ distribution and fit result for the combined mode is
shown in Fig.~\ref{fig:mbcfit} (d).
However, we provide only upper limits due to our inability
to distinguish $B \to \Kthreest\gamma$ from non-resonant decays.
The 90\% confidence level upper limit $N$ is calculated from the
relation
$\int^{N}_{0} \mathcal{L}(n)dn = 0.9 \int^{\infty}_{0} \mathcal{L}(n)dn$,
where $\mathcal{L}(n)$ is the maximum likelihood
in the $\Mbc$ fit with the signal yield
fixed at $n$. In order to include the systematic
errors from the fitting procedure
in the upper limit for the yield,
the positive systematic error is added to $N$.
The obtained yield upper limits, efficiencies and branching fractions
are listed in Table~\ref{tab:summary}.
Here, the error for the efficiency
also includes the
uncertainty in the $\Kthreest \to K\eta$ branching fraction
($(30 \pm 13)\%$).
The measurement improves the limits set by the ARGUS
collaboration~\cite{Albrecht:1988ud}.

\begin{table}
 \begin{center}
  \caption{\label{tab:summary}%
  Measured signal yields, efficiencies, branching fractions ($\BF$),
  and statistical significances (snf.)~\cite{footnote:error}.
  Efficiencies include the sub-decay branching fractions.
  Upper limits are calculated at the $90\%$ confidence level
  and include systematics.
  }
  \begin{tabular}{lcccc}
   \hline\hline
   \multicolumn{1}{c}{Mode} & Yield & Efficiency (\%)
   & $\BF$ ($\times 10^{-6}$) & snf. \\ \hline
   $B^+ \to K^+ \eta\gamma$ & $\YieldCharged$
   & $\EFFc$ & $\BFketagammaZc$ & $\SNFstatCharged$ \\
   $B^0 \to K^0 \eta\gamma$ & $\YieldNeutral$
   & $\EFFn$ & $\BFketagammaZn$ & $\SNFstatNeutral$ \\
   $B \to K\eta\gamma$ & $\YieldTotal$
   & $\EFFt$ & $\BFketagammaZt$ & $\SNFstatTotal$ \\ \hline
   $B^+ \to \Kthreest^+\gamma$ & $< \YieldULTc$
   & $\EFFTc$ & $< \BFkthreestgammaULc$ & --- \\
   $B^0 \to \Kthreest^0\gamma$ & $< \YieldULTn$
   & $\EFFTn$ & $< \BFkthreestgammaULn$ & --- \\
   $B \to \Kthreest\gamma$ & $< \YieldULTt$
   & $\EFFTt$ & $< \BFkthreestgammaULt$ & --- \\ \hline\hline
  \end{tabular}
 \end{center}
\end{table}

In conclusion, we observe the decay mode $B \to K\eta\gamma$
with a branching fraction of $\BFketagammaZtSTR$
for $\MKeta < 2.4~\GeV/c^2$.
We also set upper limits on $B \to \Kthreest\gamma$.
Although the signal yield of $B^0 \to \KS\eta\gamma$ is small,
in future this mode can be used to study
time-dependent $\CP$ asymmetry in radiative $B$ decays
and to search for new physics.


We thank the KEKB group for the excellent operation of the
accelerator, the KEK Cryogenics group for the efficient
operation of the solenoid, and the KEK computer group and
the National Institute of Informatics for valuable computing
and Super-SINET network support. We acknowledge support from
the Ministry of Education, Culture, Sports, Science, and
Technology of Japan and the Japan Society for the Promotion
of Science; the Australian Research Council and the
Australian Department of Education, Science and Training;
the National Science Foundation of China under contract
No.~10175071; the Department of Science and Technology of
India; the BK21 program of the Ministry of Education of
Korea and the CHEP SRC program of the Korea Science and
Engineering Foundation; the Polish State Committee for
Scientific Research under contract No.~2P03B 01324; the
Ministry of Science and Technology of the Russian
Federation; the Ministry of Education, Science and Sport of
the Republic of Slovenia; the National Science Council and
the Ministry of Education of Taiwan; and the U.S.\
Department of Energy.



\begin{thebibliography}{99}
\bibitem{footnote:charge}
	 Throughout this paper,
	 the inclusion of the charge conjugate mode decay is implied.
\bibitem{footnote:error}
	 The first and second errors are statistical
	 and systematic, respectively.
\end{thebibliography}
\end{document}